\documentclass[doublecol]{epl2} 

\usepackage{graphicx} 
\usepackage{epsfig} 
\usepackage{float}
\usepackage{subfig}
\usepackage[T1]{fontenc}
\usepackage[french,british]{babel}
\usepackage{amssymb}
\usepackage{epstopdf}
\usepackage{amsmath}
\usepackage{wrapfig}
\usepackage{color}
\title{Surface-wave Doppler velocimetry in a liquid metal:  \\
inferring the bifurcations of the subsurface flow.}
\shorttitle{Surface-wave Doppler velocimetry} 

\author{T. Humbert\inst{1,2} \and S. Auma\^itre\inst{2,3} \and B. Gallet\inst{2}}
\shortauthor{T. Humbert \etal}

\institute{                    
  \inst{1} Laboratoire d'Acoustique de l'Universit\'e du Maine, CNRS UMR 6613 / Univ. du Mans\\ F-72085 Le Mans Cedex 9, France\\
  \inst{2} Service de Physique de l'Etat Condens\'e, CEA, CNRS UMR 3680,
Univ. Paris-Saclay, CEA Saclay\\ 91191 Gif-sur-Yvette, France\\
  \inst{3} Laboratoire de Physique, Ecole Normale Sup\'erieure de Lyon, CNRS UMR 5672, Universit\'e de Lyon\\
 46 All\'ee d'Italie, F-69364 Lyon, cedex 07, France
}


\abstract{We introduce a velocimetry technique based on the Doppler-shift of surface waves propagating between an emitter and a receiver. In the limit of scale separation between the wavelength and the scale of the flow, we derive the direct connection between the subsurface flow and the measured phase shift between emitter and receiver. Because of its ease of implementation and high sensitivity, this method is useful to detect small velocities in liquid metal flows, where acoustical and optical methods remain challenging. As an example, we study the Hopf bifurcation of a Kolmogorov flow of Galinstan.}

\begin{document}

\maketitle

\section{Introduction}

Velocimetry is a challenging aspect of the fluid dynamics of liquid metals \cite{Shercliff,Stefani}. They are opaque, making standard optical methods impossible in the bulk of the flow. Particle-tracking velocimetry at the fluid's surface remains restricted to velocities large enough to prevent particle agglomeration \cite{Herault}, and it cannot be used when the surface is partially polluted. Acoustic Doppler Velocimetry remains a method of choice, although it is also restricted by the agglomeration and sedimentation of scatterers at low velocity. Finally, Vives probes measure the electromotive force induced by the fluid velocity near an externally imposed magnetic field \cite{Vives}. High sensitivity can be achieved for a large sensor in a strong magnetic field, but such a measurement then strongly perturbs the slow liquid metal flow.

In this Letter, we introduce an alternative method based on the Doppler-shift of surface waves by the flow. Combining several measurements of the wave field offers a way to infer the properties of the mean flow. We have implemented such a surface-wave velocimetry method in a free-surface liquid metal experiment: we detect the Hopf bifurcation of the background flow by studying the Doppler shift of surface waves propagating between an emitter and a receiver. This method displays remarkable sensitivity considering its ease of implementation, and allows us to probe the subsurface flow even when partial oxidation or surface pollution prevents horizontal fluid motion at the liquid metal's surface.

\section{Experimental set-up}
\begin{figure}
\centering{\includegraphics[width=7.5cm]{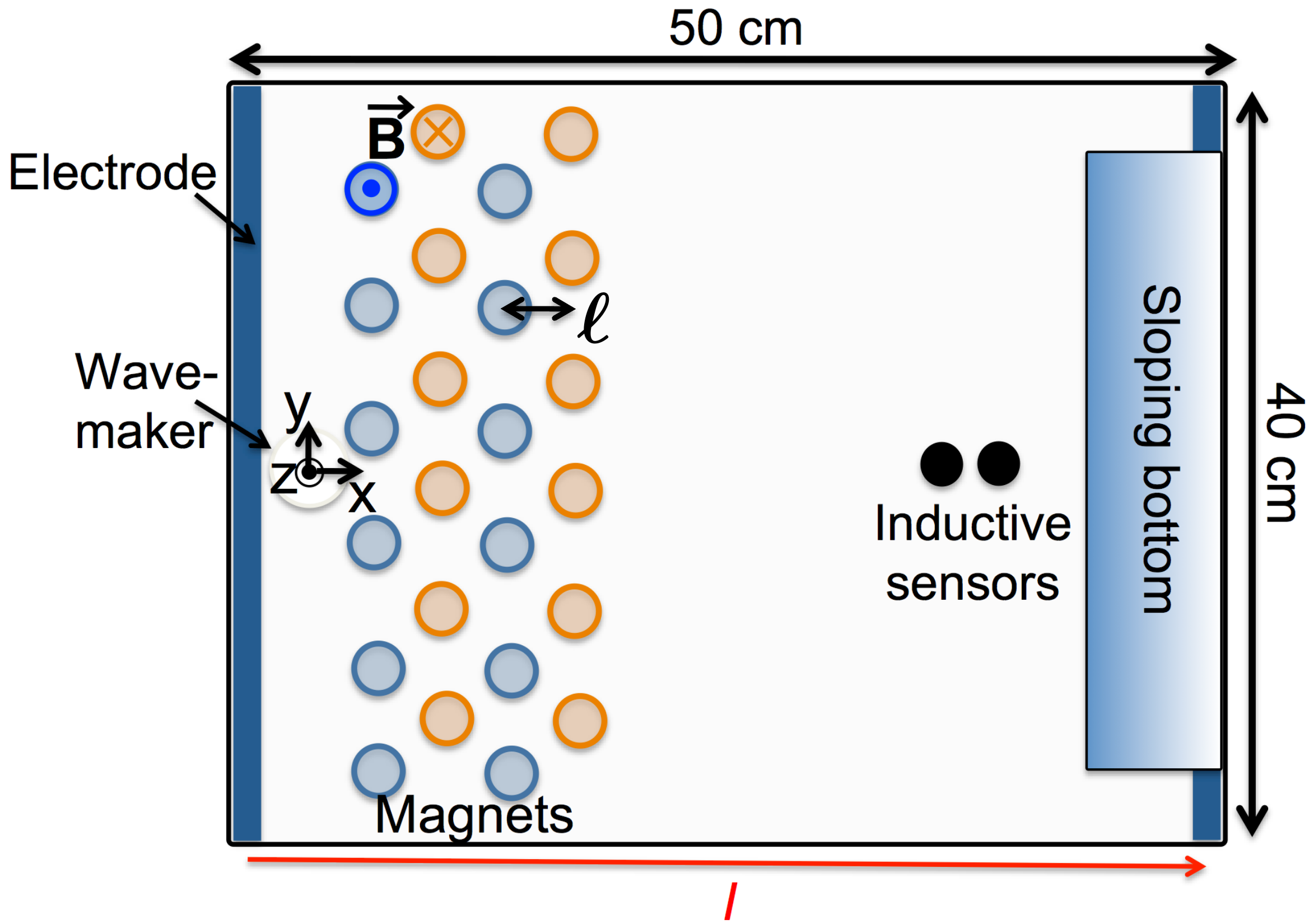}}
\caption{Top-view of the experimental cell, containing $H=1$ cm of Galinstan below $8$ cm of acidified water. Below the cell are four parallel stripes of permanent magnets. The polarity of the magnets is directed along the vertical and alternates between neighboring stripes. A current $I$ runs along $x$ between two electrodes placed at the boundaries. Its interaction with the magnetic field drives the flow through the action of the Lorentz force. Surface waves are generated near one end of the tank and measured at the other end using inductive sensors. \label{fig:setup}}
\end{figure}

\subsection{The flow}

The device generating the flow is a variation over a previously described setup \cite{Gutierrez2016, Gutierrez2016_3}. It is depicted in figure~\ref{fig:setup} and consists of a rectangular tank of length $50$ cm and width $40$ cm. It contains $H=1$ cm of Galinstan, below a layer of $8$ cm of slightly acidified water that limits the oxidation of the Galinstan surface. Galinstan is a liquid alloy at room temperature made of gallium, indium and tin. It has a density  $\rho = 6440$ kg.m$^{-3}$ and a kinematic viscosity $\nu=3.73\times10^{-7}$ m$^2.$s$^{-1}$. The fluid is stirred electromagnetically: four parallel stripes of permanent magnets are placed below the tank. They are aligned along $y$ and distant of $\ell=31.3$ mm, the dipolar axis of the magnets being along the vertical direction. The magnets of a given stripe have the same polarity, and magnets in neighboring stripes have opposite polarities (see figure~\ref{fig:setup}). The magnetic field strength at the bottom of the tank and at the center of a magnet is $B_0=0.10$ T. The combination of the four stripes then mimicks a magnetic field that oscillates in $x$. A current $I$ runs inside the liquid metal, between two electrodes placed at the far ends of the tank in the $x$ direction. Through the action of the Lorentz force, the interaction of the current with the magnetic field drives the fluid motion. Above the stripes of magnets, the Lorentz force is oscillatory in $x$ with a spatial period of $2 \ell$, mimicking a Kolmogorov forcing in this region of the tank. 

\begin{figure*}[t]
\centering{\includegraphics[height=4cm]{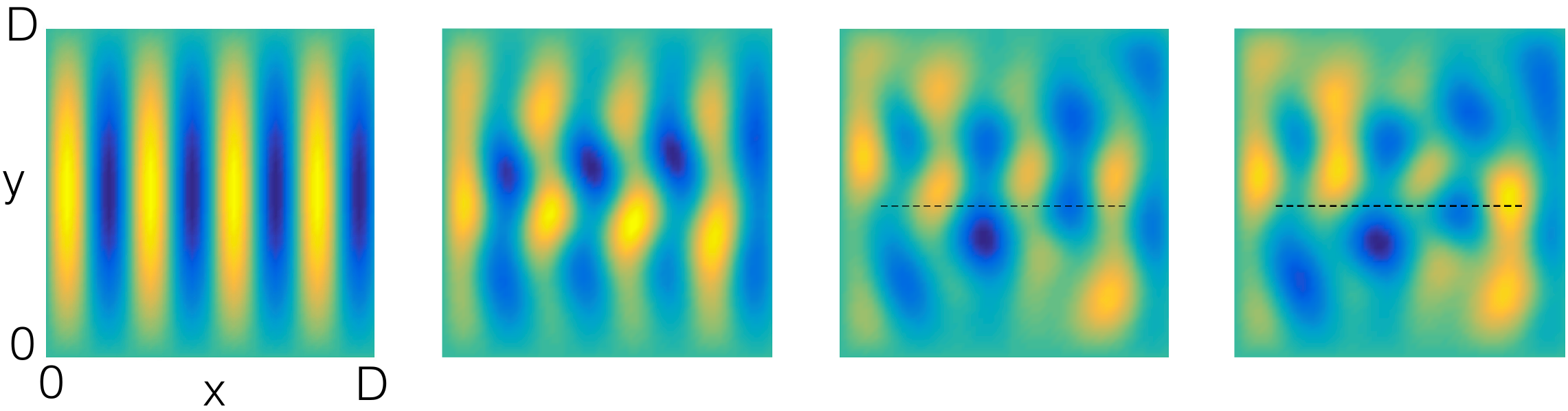}} 
          \caption{Streamfunction from numerical solutions of the 2D Navier-Stokes equation. Light color is positive and dark color is negative. Increasing the forcing from left to right: laminar solution, steady bifurcation, and oscillatory state shown at two instants separated by half a period. The dashed-line on the two last panels symbolizes the wave
path over which the velocity is integrated (from left to right) in the present velocimetry method. This integrated velocity oscillates in time: its minimum and maximum values correspond respectively to the flows displayed in the third and fourth snapshots.  \label{fig:snapshots}}
\end{figure*}

When the intensity $I$ is weak, the flow is laminar, with a geometry similar to that of the Lorentz force. As $I$ increases, this laminar solution becomes unstable. The first instabilities of such a Kolmogorov flow are well documented  \cite{Bodarenko1979,Thess1992,Juttner1997,Michel,Tithof2017}. In fully periodic geometry, and considering that the system can be approximated as a 2D one, the parallel Kolmogorov flow first undergoes a stationary bifurcation, through which it acquires  some dependence along $y$. As $I$ further increases, a Hopf bifurcation leads to a time-dependent flow. To illustrate these various regimes, we have performed numerical simulations of the 2D Navier-Stokes equation inside a square domain $[0,D]^2$ with stress-free boundary conditions, where we force a single Fourier mode $\sin(8\pi x/D)\sin(\pi y/D)$. The motion is damped through standard viscosity $\nu$ and linear friction with a friction coefficient $\mu$ satisfying $\mu=50 \pi^2 \nu/D^2$. In figure \ref{fig:snapshots} we show the successive regimes observed as we increase the forcing: laminar base-flow, steady bifurcation and Hopf bifurcation. The finite-size effects inherent to these simulations and to any experimental tank have been carefully described in Ref. \cite{Tithof2017}: because the system is not invariant to a translation along $y$ anymore, the first stationary bifurcation becomes imperfect, with no well-defined threshold value. By contrast, the Hopf bifurcation remains a ``perfect'' one, because it breaks time-invariance, which is a true symmetry of the base flow. In the following we therefore focus on this Hopf bifurcation.

\subsection{The waves}

\begin{figure}
\includegraphics[width=7cm,height=5.4cm]{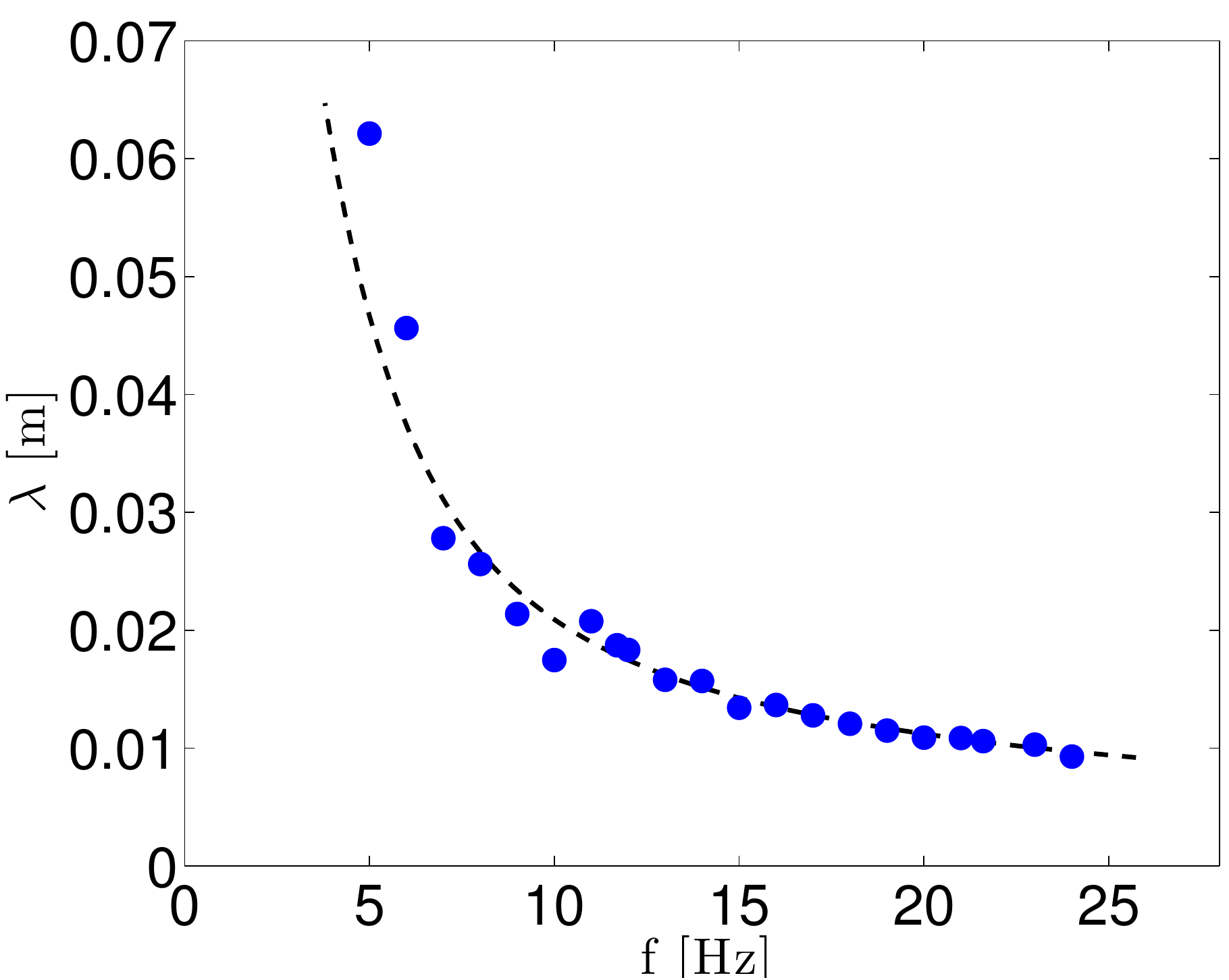}
\caption{Dispersion relation without mean flow. The blue points correspond to experimental measurements. The dashed line is the theoretical expression \eqref{reldisp} with $\sigma = 0.5$ N.m$^{-1}$. \label{fig:disprel}}
\end{figure}

Before using them to perform velocimetry, we characterize the waves on the Galinstan-water interface in the absence of mean flow. The wavemaker consists in a PVC disk of diameter 38 mm and thickness 8 mm touching the interface. An electromagnetic shaker drives a vertical sinusoidal displacement of the disk. The latter pushes the interface up and down, generating weak waves, of amplitude typically slightly below $1$ mm.  On the other side of the tank, inductive sensors mounted close to the Galinstan-water interface measure the local wave amplitude. These sensors are contactless: they are located a few millimeters above the interface and sense its height remotely. Behind the sensors, a sloping bottom ensures that the incoming waves are damped without reflection (see figure \ref{fig:setup}).

In this section we use two inductive sensors. The wavemaker and the two sensors are aligned along the $x$ axis, the distance between the two sensors being  $l_s=8.5$ mm. When the waves are on, we extract the phase difference $\theta$ between the signals of the two sensors. Provided that $l_s$ is shorter than the wavelength $\lambda$, $\theta$ is linked to the wavenumber $k$ by the simple relation $\theta=k l_s$, or equivalently:
\begin{equation}
\lambda=\frac{2\pi l_s}{\theta} \, .
\label{lambda}
\end{equation}
We apply this method to determine the wavelength over the frequency range $f \in [5, 25]$ Hz. The corresponding dispersion relation curve is displayed in figure \ref{fig:disprel}. These data can be compared to the theoretical dispersion relation of gravito-capillary waves:
\begin{equation}
\Omega(k)^2=\left( \frac{\rho-\rho'}{\rho+\rho'}gk+\frac{\sigma}{\rho+\rho'}k^3 \right) \, \tanh(kH) \, , \label{reldisp}
\end{equation}
where $\rho' = 1.0  \cdot 10^3$ kg/m$^3$ is the density of water, $g$ is the acceleration of gravity, and $\sigma$ is the surface tension between Galinstan and water. The latter was determined independently based on measurements of the Faraday instability, leading to $\sigma=0.5$ N/m~\cite{Gutierrez2016}. The theoretical expression (\ref{reldisp}) is shown as a dashed line in figure \ref{fig:disprel} and agrees well with the experimental data. In the following we therefore use (\ref{reldisp}) to extract the group velocity of the waves $v_g=\Omega'(k)$, where the prime denotes a derivative with respect to $k$. We finally stress the fact that the finite-depth correction in (\ref{reldisp}) plays a role only for the lowest frequencies visible in figure \ref{fig:disprel}. For the frequencies considered in the following, the factor $\tanh(kH)$ equals unity within $10$\% and the waves can be considered as deep-water ones.

\section{Surface-wave Doppler velocimetry}

After approximately one day, the interface between the Galinstan and the acidic solution remains clean and deformable, but it appears as a small membrane that hinders any tangential fluid motion at the interface. The properties of this interface then remain stable for weeks before further oxidation proceeds. We focus on this regime, which allows for clean reproducible measurements over several consecutive days. The goal is to detect the bifurcations of the background flow, even though no optical measurements are possible (not even surface particle tracking velocimetry). In the following we show that the Doppler-shift induced by the subsurface flow on the waves can be used as a mean to probe the flow.

We consider only one of the two inductive sensors shown in figure \ref{fig:setup}, and we also measure the displacement of the wavemaker. The relative phase difference between the wave amplitude measured at large $x$ and the displacement of the wavemaker is denoted as $\phi$. We extract it using standard Hilbert transform of the two oscillatory signals. We denote as $\Delta \phi \, (t)$ the difference in $\phi$ when the mean-flow is on ($I \neq 0$) and off ($I=0$):
\begin{equation}
\Delta \phi (t)= \phi(t)-\phi|_{I=0} \, .
\end{equation}

\begin{figure}[t]
   \centering{\includegraphics[width=7.3cm,height=5.5cm]{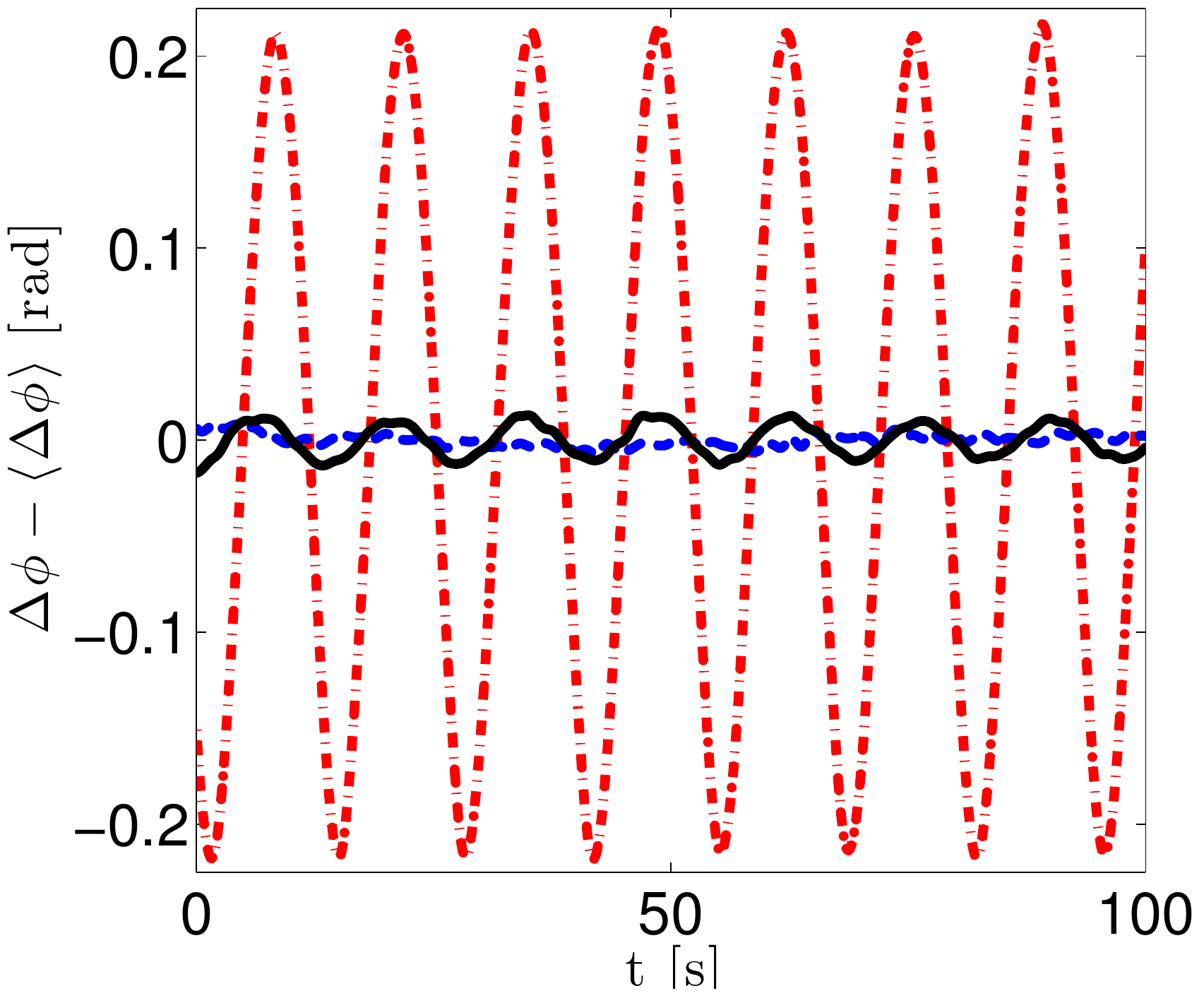}}
          \caption{Phase shift $\Delta \phi \, (t)-\langle\Delta \phi \, (t)\rangle$ as a function of time, where $\langle \cdot \rangle$ denotes time average, for $f=21.6$ Hz. \textcolor{blue}{- -}: $I=2A$, \textcolor{black}{-}: $I=2.12$ A, \textcolor{red}{-.}: $I=2.5$ A.
    \label{fig:time_signal}}
\end{figure}

The fluctuating part of this phase difference is shown in figure \ref{fig:time_signal} for several values of the current $I$. For $I \leq 2$ A, the phase shift is independent of time. By contrast, for greater values of $I$, $\Delta\phi$ depends on time: it is oscillatory, with a period greater than ten seconds and an amplitude that increases with $I$. We interpret this phenomenon as a consequence of a Hopf bifurcation of the background Kolmogorov flow: the unstable oscillatory mode leads to a nonzero average fluid velocity along $x$ over the distance of propagation. The cumulative effect of this advecting velocity Doppler-shifts the waves, therefore changing the phase shift $\phi(t)$ between the wavemaker and the inductive sensor. The oscillations in $\Delta \phi \, (t)$ therefore originate from the oscillations of the background mean-flow.

The wave amplitude is below $0.5$ mm  near the wavemaker, and it is much smaller near the remote inductive sensor because of both viscous damping and radial spreading of the wave energy around the wavemaker. A key aspect of the measurement technique is that it is insensitive to such damping and radial spreading of the wave energy, because the information on the background flow velocity is encoded in the phase of the wave and not in its amplitude.
We also made sure that the waves do not affect the mean flow, by checking that the signals in figure \ref{fig:time_signal} are rather insensitive to the wave amplitude: at most, the amplitude of oscillation of $\Delta \phi$ decreases by 20\% when the wave amplitude is multiplied by six.

\section{Bifurcation curves}
The present Hopf bifurcation cannot be directly compared to previous experiments or analytical results on Kolmogorov flows, because the lines of circular permanent magnets somewhat differ from the standard experimental and theoretical setups. The closest experiment may be the one of Michel et al. \cite{Michel}. In such shallow liquid metal experiments, the main damping mechanism is Hartmann friction near the bottom boundary. In dimensionless form, this phenomenon is quantified by the parameter $Rh=\sqrt{I H /(B_0 \nu \sigma L^2)}$ (see e.g. Ref. \cite{Sommeria} for details). At the bifurcation threshold, using $B_0$ for the typical magnetic field amplitude, we obtain $Rh=1.0$: it lies within a factor of two of the threshold value of the Hopf bifurcation reported by Michel et al. \cite{Michel}, which is acceptable considering the fact that the structure of the forcing strongly differs in the two experiments. In the remainder of this section, we characterize in more details the Hopf bifurcation observed in the present experiment.

\begin{figure}[t]
   \centering{\includegraphics[width=7.1cm,height=5.5cm]{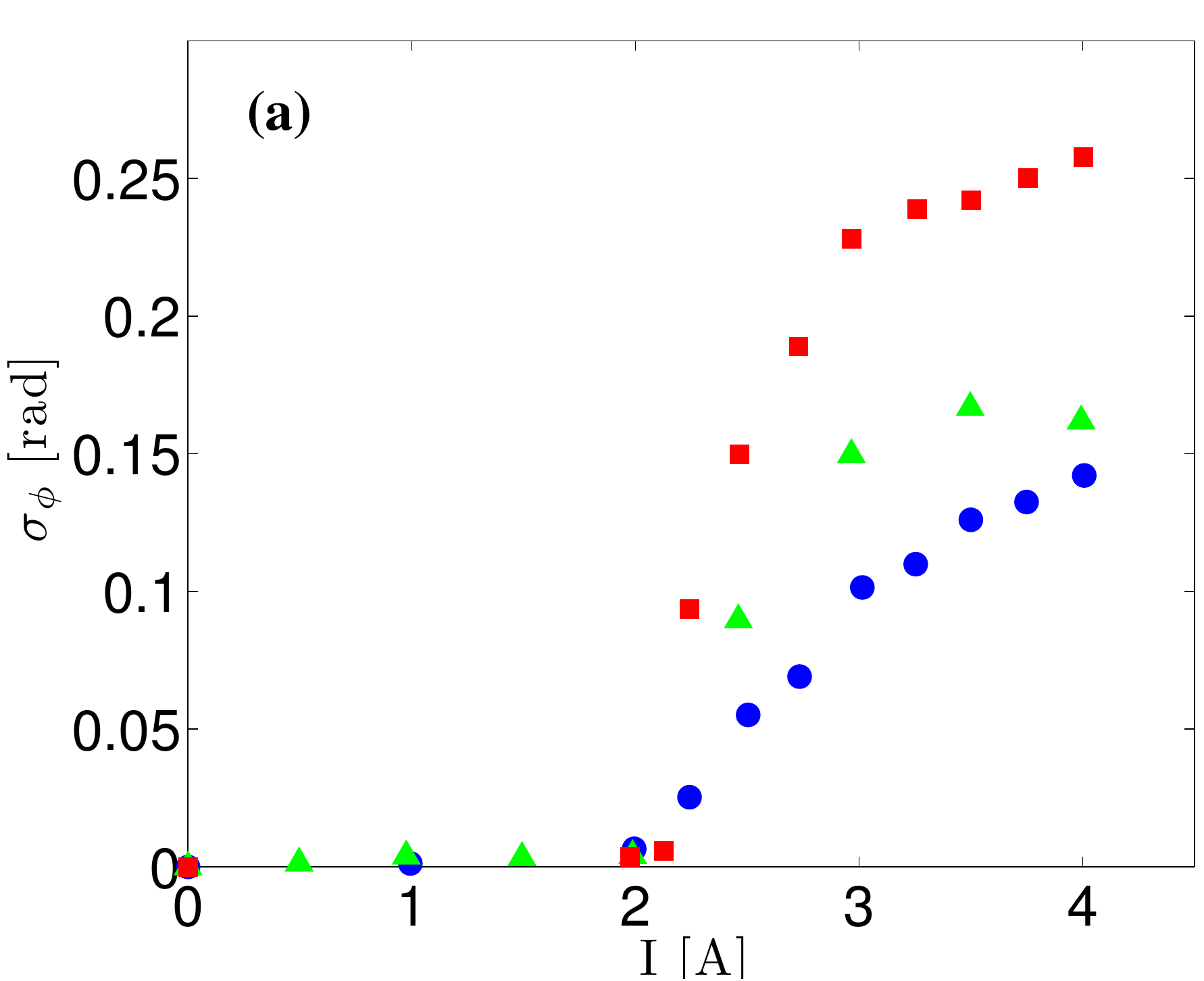}}\\
\centering{\includegraphics[width=7cm,height=5.7cm]{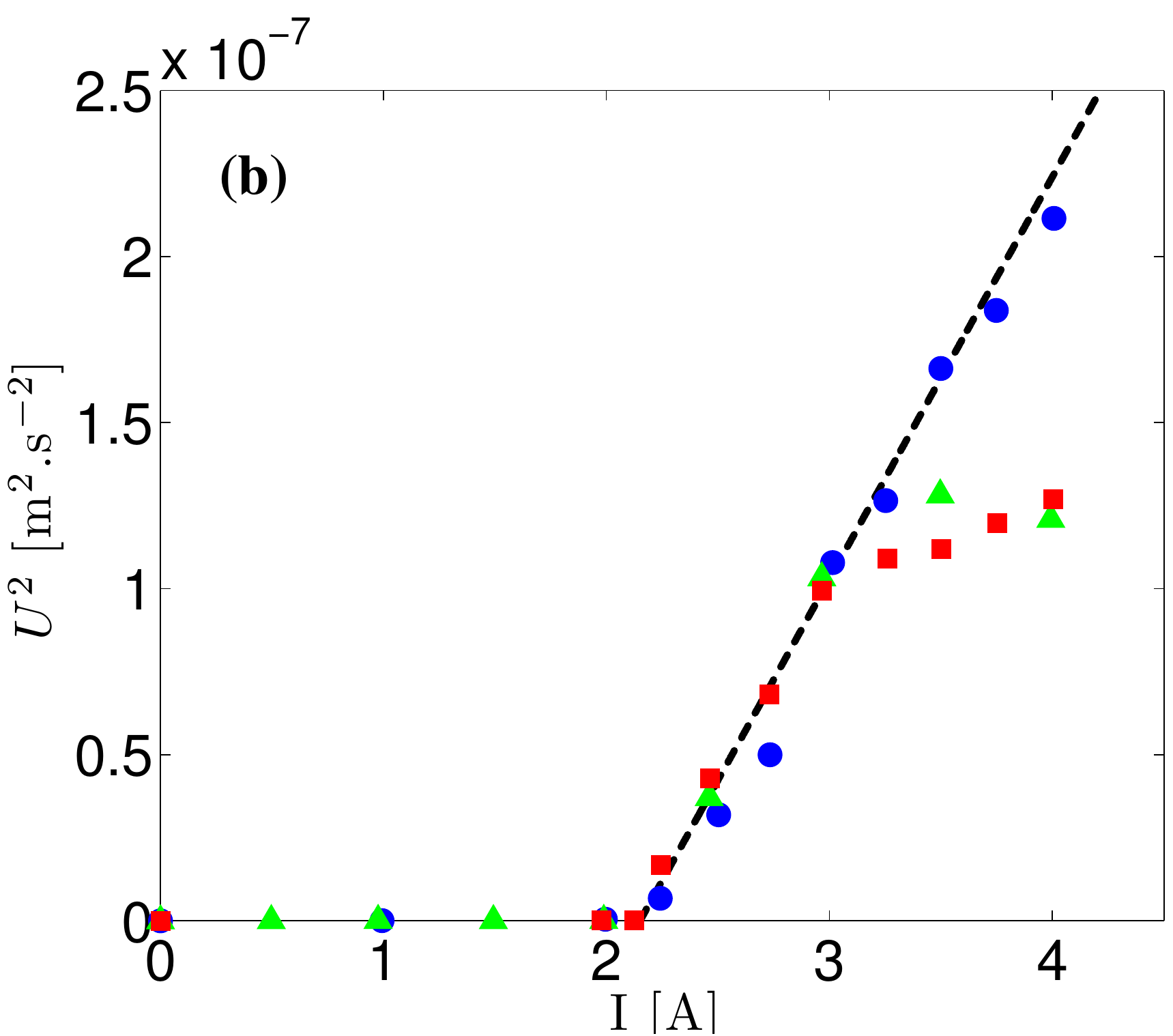}}\\
\centering{\hspace{0.1cm}\includegraphics[width=7.3cm,height=5.5cm]{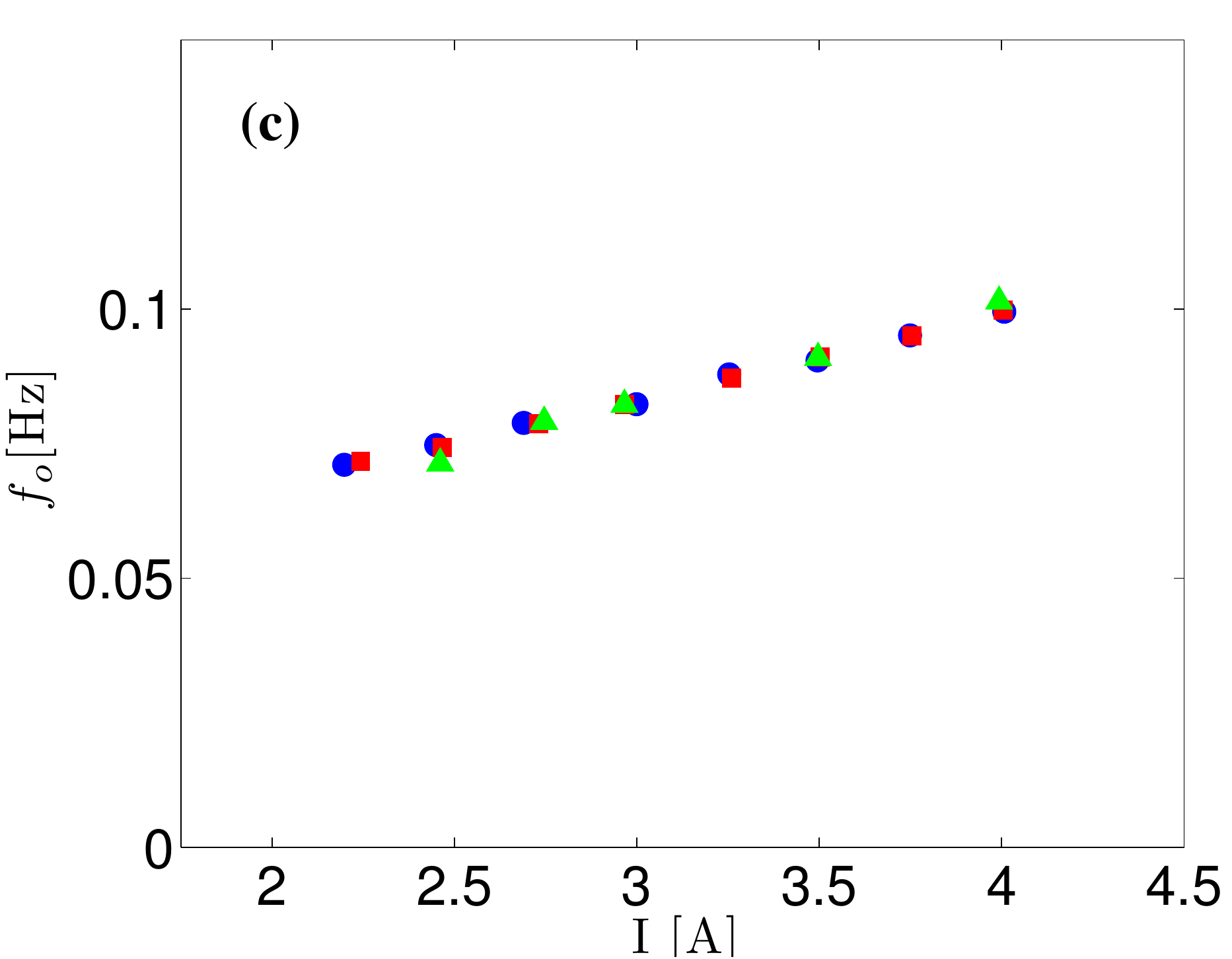}}
          \caption{(a) Standard deviation of the phase difference $\Delta\phi \, (t)$ as a function of the input current $I$. (b) The square of the typical velocity $U$ defined in (\ref{defU}) as a function of $I$. (c) Frequency $f_o$ of the oscillations of $\Delta\phi \, (t)$ as a function of $I$. Blue circles: $f=6$ Hz. Green triangles: $f=11$ Hz. Red squares: $f=21.7$ Hz.
    \label{fig:std_phi}}
\end{figure}

From the time series displayed in figure \ref{fig:time_signal} we extract the root-mean-square value $\sigma_\phi$ of $\Delta \phi \, (t)$ and plot it as a function of $I$ in figure \ref{fig:std_phi}a. The oscillation appears above a threshold value of the current $I$, i.e., above a threshold velocity of the base flow. $\sigma_\phi$ evolves as the square root of the departure from onset, in agreement with the standard phenomenology of Hopf bifurcations. We show in panel \ref{fig:std_phi}c the frequency of the oscillatory eigenmode, which is also the frequency $f_o$ at which the phase-difference oscillates. Of course, waves at different frequencies lead to the same oscillation frequency $f_o$, because they sense the same oscillatory eigenmode of the subsurface flow. However, they are not Doppler-shifted by the same phase difference, and the measured value of $\sigma_\phi$ depends on the frequency of the surface waves. In the following we elaborate on the precise relationship between the phase-shift $\Delta \phi$ and the subsurface flow.

\section{Interpretation}

The relation between the phase difference $\Delta \phi \, (t)$ and the subsurface flow can be made more quantitative in the limit of scale separation. 
For a weak depth-dependent flow $u(z)$ directed along $x$, a standard asymptotic expansion allows one to write the dispersion relation of deep-water surface waves propagating along $x$ as:
\begin{equation}
\omega= \Omega(k) + \tilde{u} \, k \, , \label{Dshift}
\end{equation}
where $\omega$ is the wave angular frequency, the function $\Omega(k)$ is defined in (\ref{reldisp}), and only the first correction in $u/v_g$ is retained. The latter correction involves a weighted vertical average of the velocity profile \cite{Stewart1974}:
\begin{equation}
\tilde{u} = 2 k \int_{-\infty}^0 e^{2k z} u(z) \, \mathrm{d}z \, , \label{zavg}
\end{equation}
this expression being valid for a semi-infinite fluid layer occupying $z<0$ initially. This approximation holds as long as $e^{-2k H} \ll 1$, which is satisfied in the present experiment. We stress the fact that even when partial oxidation prevents horizontal motions directly at the liquid metal surface, the subsurface flow nevertheless leads to a nonzero value of $\tilde{u}$ in (\ref{zavg}) and therefore to a Doppler-shift of the surface waves.

The dispersion relation (\ref{Dshift}) can be readily extended to a flow ${\bf u}(x,y,z)$ that depends slowly on $x$ and $y$ as compared to the wavelength. We obtain a relation between the frequency and the local wavenumber $k(x)$ of the waves:
\begin{equation}
\omega= \Omega[k(x)] + \tilde{u}(x) \, k(x) \, , \label{dispcurrent}
\end{equation}
where $u$ denotes the $x$-component of ${\bf u}$ and the $\tilde{\cdot}$ denotes the weighted vertical average (\ref{zavg}).
In the present setup the wave frequency is imposed by the wavemaker and we write $\omega=\omega_0=\text{const}$. By contrast, the local wavenumber $k(x)$ adapts to the local fluid velocity. We denote as $k_0$ this wave number in the absence of mean-flow, such that $\omega_0=\Omega(k_0)$. For weak mean-flows, $k(x)$ departs only slightly from $k_0$ and we can expand  (\ref{dispcurrent}) to first order in $k(x)-k_0$, which yields:
\begin{equation}
\omega_0= \Omega(k_0) + \Omega'(k_0)\,[k(x)-k_0] + \tilde{u}(x) \, k_0 \, . 
\end{equation}
Denoting as $v_g=\Omega'(k_0)$ the group velocity of the waves in the absence of currents, we obtain the local wavenumber:
\begin{equation}
k(x)= k_0 \left( 1 - \frac{\tilde{u}(x)}{v_g} \right) \, .
\end{equation}
The wavenumber being the local gradient of the phase, we integrate this relation to get:
\begin{equation}
\phi=k_0 \int_{x=0}^{x=L} \left[1 - \frac{\tilde{u}(x)}{v_g} \right] \, \mathrm{d}x \, ,
\end{equation}
where $x=0$ denotes the position of the wavemaker and $x=L=0.34$ m the position of the inductive sensor. Subtracting the same expression for $\tilde{u}=0$, we finally obtain:
\begin{equation}
\Delta \phi= - k_0 \int_{x=0}^{x=L} \frac{\tilde{u}(x)}{v_g}  \, \mathrm{d}x \, . \label{eqth}
\end{equation}
Because the travel time of the waves between emitter and receiver is much faster than the timescale of the mean flow evolution, we can use equation (\ref{eqth}) at any time to compute the phase shift $\Delta \phi \, (t)$ due to the slowly evolving mean-flow, substituting in (\ref{eqth}) the instantaneous velocity $\tilde{u}(x,t)$.

Equation (\ref{eqth}) provides a direct connection between the phase difference and the subsurface flow: we estimate $\tilde{u}$ by defining the typical sensed velocity as:
\begin{equation}
U=\frac{v_g \lambda \sigma_\phi}{2 \pi L} \, . \label{defU}
\end{equation}
In figure \ref{fig:std_phi}b, we plot $U^2$ as a function of $I$. In the vicinity of the bifurcation threshold, this representation leads to a good collapse of the data obtained for different wave frequencies, which indicates that U defined in (\ref{defU}) is indeed the right order parameter. This collapse also confirms that the wave amplitude is irrelevant provided one stays in the linear regime: the weak waves do not affect the background flow. The collapse remains satisfactory even for the lowest wave frequency, $f=6$ Hz, for which there is no real scale separation between the wavelength $\lambda$ and the typical scale $2 \ell$ of the flow. This may come from the fact that the small-scale velocity fluctuations (visible in the numerical snapshots \ref{fig:snapshots}) cancel each other in expression (\ref{eqth}). Only the larger-scale velocity structures leading to a nonzero cumulative average when integrated over $x$ contribute to (\ref{eqth}), and these structures indeed have a large typical scale as compared to $\lambda$.

Equation (\ref{defU}) also provides an order of magnitude of the sensitivity of the surface-wave Doppler velocimetry method. In figure \ref{fig:time_signal}, we show a rather clean oscillation signal measured in the immediate vicinity of the bifurcation threshold ($I=2.12$ A). The corresponding typical velocity is $U \simeq 1 \cdot 10^{-5}$m.s$^{-1}$. Although unknown geometrical prefactors may affect the relationship between $U$ and the true velocity, this low value indicates a remarkable sensitivity of the measurement technique, making it competitive as compared to available velocimetry methods for the low-velocity regime of liquid metal hydrodynamics.

\section{Conclusion}
We have presented an original velocimetry technique where the Doppler-shift of surface waves is used to detect the subsurface flow. It is particularly well-suited to detect small velocities in liquid metals, where partial oxidation and pollution of the surface prevent the use of standard optical methods such as surface particle-tracking velocimetry. As an example of such an application, we precisely measured the first Hopf bifurcation of a Kolmogorov flow of Galinstan. 
In the limit of scale separation between the wavelength and the lengthscale of the flow, we established the direct connection between the subsurface flow and the measured phase shift. This led to a good collapse of the data obtained at various wave frequencies.
Another possible application of this method would be to detect mean flows under complex surface-wave fields. Indeed, surface particle tracking is then strongly perturbed by the Stokes drift of the surface waves, and PIV is difficult if the particles are to be imaged through the deformed free-surface. If the waves are linear, one can externally superimpose to the complex wave field a well-controlled monochromatic wave and detect the mean-flow through a lock-in detection of the phase difference between the emitter and a remote sensor. In any case, the advantages of the present method are its ease of implementation and accuracy: it efficiently detects weak velocity signals in free surface flows. Of course, the method also has limitations, one of which being that the measurement integrates the velocity along the path between the wave emitter and receiver, leading to a rather nonlocal measurement. An exciting next step would be to combine several emitters and receivers to image the mean-flow through surface-wave tomography.

\acknowledgments
We thank V. Padilla for building the experimental setup. This research is supported by Labex PALM ANR-10-LABX-0039 and ANR Turbulon 12-BS04-0005.


\begin{thebibliography}{0}


\bibitem{Shercliff}
\Name{\sc Shercliff J. A.} 
\Book {The theory of electromagnetic flow-measurement} 
\Publ{Cambridge University Press}
\Year{1987}

\bibitem{Stefani}
\Name{\sc Eckert S., Buchenau  D., Gerbeth G., Stefani F. \and  Wiess F.-P.}
\REVIEW{Journal of Nuclear Science and Technology}{48}{2011}{490-498}.

\bibitem{Herault}
\Name{Herault J., P\'etr\'elis F. and Fauve S.}
\REVIEW{Europhysics Letters }{111}{2015}{44002}.

\bibitem{Vives}
\Name{\sc Ricou R. \and  Vives C.}
\REVIEW{International Journal of Heat and Mass Transfer}{25}{1982}{1579-1588}.



\bibitem{Gutierrez2016_3}
\Name{\sc Guti\'errez P. \and Auma\^itre S.}
\REVIEW{European Journal of Mechanics-B/Fluids}{60}{2016}{24-32}.


\bibitem{Gutierrez2016}
\Name{\sc Guti\'errez P. \and Auma\^itre S.}
\REVIEW{Physics of Fluids}{28}{2016}{025107}.

\bibitem{Bodarenko1979}
\Name{\sc Bodarenko N. F., Gak M. Z. \and Dolzhansky F. V.}
\REVIEW{Izv. Akad. Nauk (Fiz. Atmos. Okeana)}{15}{1979}{1017}.

\bibitem{Thess1992}
\Name{\sc Thess A.}
\REVIEW{Physics of Fluids A}{4}{1992}{1385-1395}.

\bibitem{Juttner1997}
\Name{\sc J{\"u}ttner, B., Marteau, D. Tabeling, P. \and Thess, A}
\REVIEW{Physical Review E}{55}{1997}{5479-5488}.

\bibitem{Michel}
\Name{Michel, G., Herault J., P\'etr\'elis F. and Fauve S.}
\REVIEW{Europhysics Letters }{115}{2016}{64004}.

\bibitem{Sommeria}
\Name{\sc Sommeria J.}
\REVIEW{Journal of Fluid Mechanics}{170}{1986}{139-168}.


\bibitem{Tithof2017}
\Name{\sc Tithof J., Suri B., Pallantla R. K., Grigoriev R. O. \and Schatz M. F.}
\REVIEW{Journal of Fluid Mechanics}{828}{2017}{837-866}.

\bibitem{Stewart1974}
\Name{\sc Stewart, R. H. \and Joy, J. W.}
\REVIEW{Deep Sea Research and Oceanographic Abstracts}{21}{1974}{1039-1049}.





\end{thebibliography}
\end{document}